\documentclass[aps,prl,reprint,groupedaddress]{revtex4-2}
\usepackage[english]{babel}
\usepackage[printonlyused]{acronym}
\usepackage{float}
\usepackage[hidelinks]{hyperref}
\usepackage{graphicx}
\usepackage{subcaption}
\usepackage{dcolumn}
\usepackage{bm}
\usepackage{makecell}
\usepackage{array}
\usepackage{acronym}
\usepackage{rotating}
\usepackage{hyperref}
\usepackage{orcidlink}
\usepackage{tabularx}
\usepackage{booktabs}
\usepackage{braket}
\usepackage{bm}
\usepackage{framed}
\hypersetup{breaklinks=true}
\usepackage[english]{babel}
\usepackage{caption}
\usepackage{mdframed}
\usepackage{enumitem}
\begin{document}

\title{Comparing Visual Qubit Representations in Quantum Education: The Bloch sphere Enhances Task Efficiency}

\author{Linda Qerimi}
\email{linda.qerimi@physik.uni-muenchen.de}
\affiliation{Chair of Physics Education, Faculty of Physics, Ludwig-Maximilians-Universität München, Geschwister-Scholl-Platz 1, 80539 Munich, Germany, and \\Munich Quantum Valley (MQV), Max Planck Institute of Quantum Optics (MPQ), Germany}

\author{Sarah Malone}
\affiliation{Department of Education, Saarland University,  Campus A4 2, 66123 Saarbrücken, Germany}

\author{Eva Rexigel}
\affiliation{Department of Physics and State Research Center OPTIMAS, University of Kaiserslautern-Landau, Erwin-Schroedinger-Str. 46, 67663 Kaiserslautern, Germany}

 \author{Jochen Kuhn}
\affiliation{Chair of Physics Education, Faculty of Physics, Ludwig-Maximilians-Universität München, Geschwister-Scholl-Platz 1, 80539 Munich, Germany}

\author{Stefan Küchemann}
\affiliation{Chair of Physics Education, Faculty of Physics, Ludwig-Maximilians-Universität München, Geschwister-Scholl-Platz 1, 80539 Munich, Germany}

\date{\today}

\begin{abstract}
Visual-graphical qubit representations offer a means to introduce abstract quantum concepts -- such as quantum state, superposition, or measurement -- in an accessible manner, particularly for learners with low prior knowledge. Building on a previous expert rating of the mechanisms of qubit representations, this study compared two representations -- the Bloch sphere and the Quantum Bead -- in terms of learning outcomes, task performance, cognitive load, and mid-term retention.
The study was conducted with  N=149 secondary school students. The study assessed conceptual understanding via pre- and post-tests, application-oriented task performance by measuring accuracy per time, and cognitive load via intrinsic, extrinsic and germane dimensions. A follow-up test after 1-2 weeks assessed medium-term retention.
Results showed no significant effect of representation type on post-test learning outcomes. However, process data revealed that learners using the Bloch sphere completed application-oriented tasks significantly more efficiently. The cognitive load was similar in both groups.  Mid-term retention of quantum concepts was stable across groups, and early learning performance emerged as the strongest predictor of mid-term retention.

In conclusion, these findings emphasize that instructional impact is not solely determined by outcome measures, but also by how representations influence cognitive processing, task integration, and learners’ interaction with complex content.

\end{abstract}

\maketitle

\section{\label{sec:levelA}I. Introduction}

\subsection{A. Motivation und Backgroud}

Learning in \ac{qp} has become increasingly relevant, not least because of the growing importance of emerging \ac{qt} such as quantum computing, quantum cryptography, and quantum sensing. The physical conditions under which these technologies can be realised, for example, through ion traps, superconducting or neutral atoms, and are an active subject of fundamental research(e.g.~\cite{mqv2025}).
Key concepts such as quantum measurement, superposition, probability, and entanglement are essential across all implementations. The interest in teaching these concepts now spans a broad spectrum of learners from school to industry with varying levels of prior knowledge and diverse learning goals~\cite{greinert_future_2023, merzel_core_2024}. The use of representations in QP enables learners to gain access to an otherwise highly abstract and formalised world. Particularly in a school context, where mathematical skills are often not yet fully developed, suitable representations can help to make fundamental concepts of QP understandable.
Various forms of representation are available in QP, ranging from symbolic-mathematical representations (such as Dirac notation or Schrödinger formalism) to visual-graphical representations such as the Bloch sphere. The latter offer the possibility of making central concepts such as superposition, measurement or quantum state accessible by visual means, independently of an intensively formal derivation~\cite{dur_was_2012, dur_visualization_2014}. The review by Donhauser et al. (2024) analyses teaching and learning elements, innovative tools and process-oriented studies from the period between 2018 and early 2024~\cite{PhysRevPhysEducRes.20.020601}. The results show that more than half of the articles examined are based on the qubit idea methodology or are preparing for it~\cite{PhysRevPhysEducRes.20.020601}. This demonstrates the upcoming relevance of qubits as teaching strategies~\cite{PhysRevPhysEducRes.20.020601}. In particular, external representations of two-level quantum systems -- referred to as qubits -- combined with the 'spin-first' approach, are well suited for introducing the elementary structure of QP in a clear and accessible manner and for embedding it into contextualised learning environments~\cite{sadaghiani_spin_2015, dur_visualization_2014, muller_milqquantum_2021}. Visual representations can establish a connection between phenomena and concepts, thereby supporting conceptual development.
In QP, it is more difficult to find consistent visualisations than for classical physics lessons~\cite{stadermann_analysis_2019}, so the selection and design of suitable representations is of particular importance. In our previous study, we identified 16 features of visual-graphical qubit representations that are generally capable of supporting students during learning with visual representations ~\cite{qerimi2025exploring}. These features were rated by experts on a scale of 1 to 5 for specific qubit representations. The results show that experts anticipate differences in the support of specific learning processes and suggest that certain representations activate different mechanisms in the learning process~\cite{qerimi2025exploring}.
The expert ratings provide insights into how qubit representations are evaluated with regard to features that support learning. However, it remains unclear how students actually perceive these representations and benefit from them. In the present study, two representations were selected based on their expert ratings to investigate whether differences in learning outcomes emerge and to what extent these align with the expert evaluations. To address this aspect, we focus on the following research questions (RQs) and research hypotheses (RHs). 

\subsection{B. Research Questions and Hypotheses}

The study examines whether visual representations support students' conceptual understanding of QP, and whether the Bloch sphere and quantum beads representations differ in their instructional effectiveness, particularly in the context of QT.\\

\noindent\textbf{RQ1:} To what extent do different visual-graphical representations (Quantum Bead vs. Bloch sphere) foster learning quantum concepts differently? \\
\textbf{H1.1:} Participants who learn with the Bloch sphere achieve a higher learning outcome than those who learn with the Quantum Bead representation.\\
\textbf{H1.2:} Participants using the Bloch sphere perform more efficient on application-oriented quantum tasks in \textit{phase gate, amplitude, quantum state, superposition, quantum measurement}, than those using the Quantum Bead.\\

\noindent\textbf{RQ2:} How do different visual-graphical representations (Bloch sphere vs. Quantum Bead) affect the use of cognitive resources in the learning of quantum concepts?
\textbf{H2:} Participants who learn with the Bloch sphere show a more effective use of cognitive resources than those who learn with the Quantum Bead.\\

\noindent\textbf{RQ3:} How does the use of different visual-graphic representations (Quantum Bead and Bloch sphere) influence medium-term retention of fundamental quantum concepts?\\
\textbf{H3:} Learners who learned the Bloch sphere will demonstrate higher medium-term retention of basic quantum concepts compared to those who use the Quantum Bead.

\section{II. Theory \label{sec:levelB}}

\subsection{A. Theoretical Foundations of Representational Learning}

Learning environments in physics -- especially in QP -- often involve a variety of representations, such as symbolic, verbal (e.g. text-based) or visual- graphical representations~\cite{lemke_multiplying_1998}. Learners are often faced with the challenge of linking these representations and integrating them conceptually. At the same time, working with multiple representations offers educational opportunities. A key finding of the meta-analysis by Rexigel et al. (2025) suggests that the advantages of \ac{mer} are not limited to established combinations of two forms of representation~\cite{rexigel2024more}. Positive effects are also evident when three or more representations are combined~\cite{rexigel2024more}.

A key task for learners is to understand the individual representations and to extract and connect essential information from them in order to form coherent mental representations (schemata)~\cite{schnotz_construction_2003, schnotz_kognitive_2001}. The cognitive processes involved in this integration are described, for example, in the Cognitive Theory of Multimedia Learning (\acf{ctml}~\cite{mayer_fiorella_2014}. The CTML is based on fundamental assumptions such as the limited capacity of working memory and posits processes such as \textit{selecting}, \textit{organizing}, and \textit{integrating} (SOI) information as central to meaningful learning~\cite{Mayer.2021e}. To support these processes and facilitate efficient learning, instructional design should aim to optimize the use of limited working memory resources, thereby promoting effective learning outcomes.

These considerations set the stage for examining the cognitive mechanisms that underlie learning with (multiple) external representations.

\subsubsection{Cognitive Theory of Multimedia Learning}

The multimedia principle states that learners understand content better when it is presented both verbally and visually than when it is purely text-based~\cite{Mayer.2021e}. Learning QP with multiple representation offers the opportunity to utilise this advantage and support students in the learning process~\cite{mayer_fiorella_2014}. 

The CTML is based on the following three assumptions: channel duality, limited capacity and active processing~\cite{Mayer.2021e}. The term ‘dual channels’ refers to different modalities, such as visual and verbal perception~\cite{Mayer.2021e}. 
\begin{figure}[h!]
  \centering
  \includegraphics[width=0.5\textwidth]{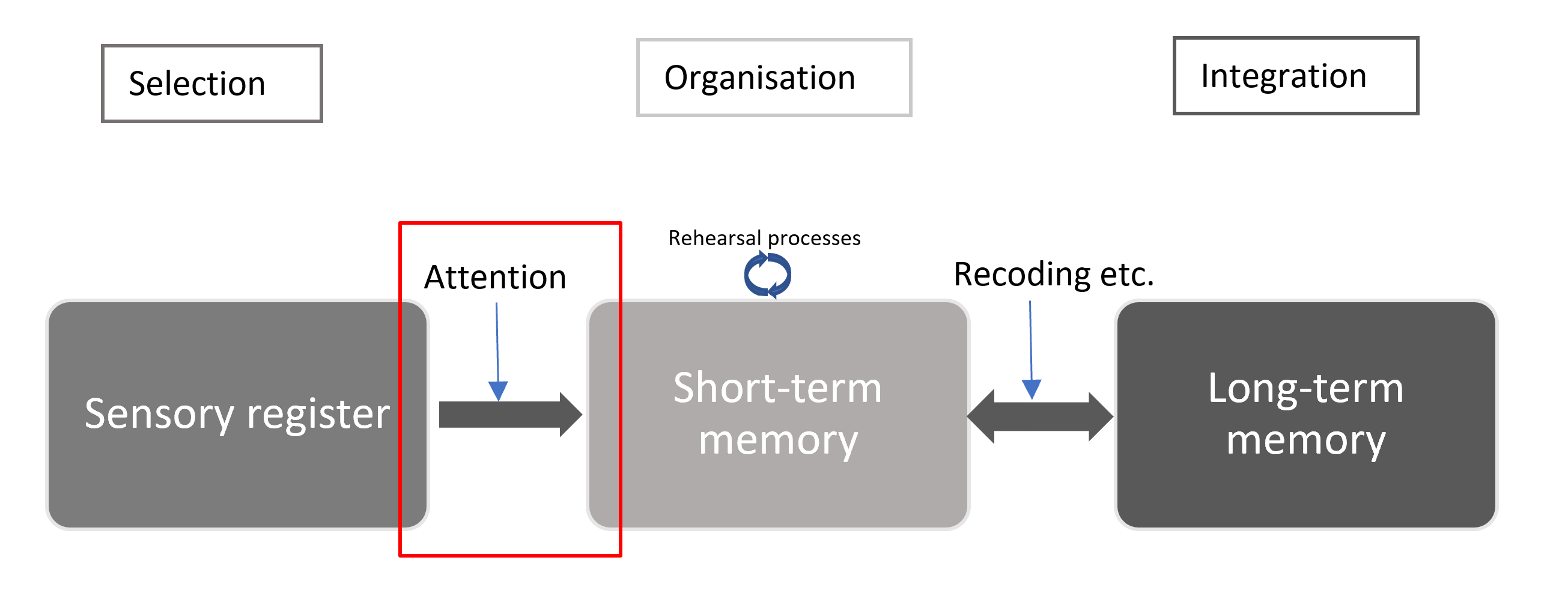} 
  \caption{Based on the multi-store model of memory, cognitive processing involves the temporary storage and manipulation of information in working memory. Through processes such as rehearsal, elaboration, and integration, information is transferred to mid-term memory and linked to prior knowledge. Effective learning requires meaningful organization and retrieval of both existing and newly acquired information.}
  \label{fig:atkinson-shiffrin}
\end{figure}

Limited capacity refers to the limited number of information in working memory that can be processed simultaneously~\cite{Mayer.2021e, sweller_cognitive_2019, sweller_cognitive_1994}. Active processing involves selecting relevant information so that it can be further processed through organisational and structural processes~\cite{Mayer.2021e}. The goal of this integration is to connect new information with existing knowledge. 

Mayer (2014) sees multimedia design as a possible approach to support learners in mental modelling~\cite{Mayer.2021e}. 

According to Mayer (2014), two implications for multimedia design can be derived from this assumption: (1) The material presented should have a coherent structure and (2) the message should provide learners with guidance on how the structure can be constructed~\cite{mayer_fiorella_2014}. 

To support this goal, multimedia design can be used strategically, in combination with knowledge and findings of representational mechanisms, to shape the design of learning environments and promote the development of a functional mental model~\cite{mayer_fiorella_2014, ainsworth_deft_2006, qerimi2025exploring}.

\subsubsection{Cognitive Load}

Any new information presented to a learner puts a strain on working memory, which refers to the short-term memory system and has a limited capacity to process and organise information (see Figure \ref{fig:atkinson-shiffrin}) \cite{mayer_fiorella_2014, sweller_cognitive_2019, sweller_cognitive_1994}. Ideally, learning processes in working memory allow not only for processing but also for initial integration with prior knowledge.
According to Sweller (1998), cognitive load can be categorised into three types~\cite{sweller_cognitive_1994}: 
\begin{itemize}
\item intrinsic cognitive load, which arises from the inherent complexity and difficulty of the learning material, 
\item extraneous cognitive load, which is caused by irrelevant or poorly designed information from outside the content, and 
\item germane cognitive load, which refers to the proportion of cognitive resources that directly support learning and the construction of mental structures. 
\end{itemize}

The interplay between new information about representation and concepts in QP can increase and strain processing capacity, thereby becoming a barrier to learning~\cite{sweller_cognitive_2019}. But Sweller (1988, 1998) also demonstrates, that highlighted elements in instructional materials can facilitate processing and thereby enhance learning~\cite{sweller_cognitive_1988, sweller_cognitive_1994}. However, visual-graphical representations have the potential to reduce cognitive load if they are accessible and stimulate learners' prior knowledge and different processing channels~\cite{mayer_fiorella_2014, baddeley_working_1992}. The \ac{clt} emphasizes that learning is most effective when instructional design takes into account the limitations of working memory\cite{sweller_cognitive_1994}. Selecting appropriate representations plays a crucial role in optimizing cognitive load -- not necessarily by reducing it, but by aligning it with the task and learners’ prior knowledge.

\subsubsection{Process-Based Indicators of Learning}

Learning success is often assessed based on learning outcomes such as test results or progress in conceptual understanding. However, learning is a process and important aspects of this process cannot always be captured directly by measuring outcomes (e.g.~\cite{huang2009measuring}).
\\
\paragraph{Time:} To gain a deeper insight into how learners deal with representations and tasks, it is essential to consider process-related indicators such as reaction time, subjective confidence or even visual attention (e.g. through eye tracking). Hou and Zhang (2006) show that visual information processing is dependent on processing time~\cite{hou2006time}. The longer a visual stimulus is viewed, the greater the resolvable depth of detail, particularly for complex or finely structured content. The authors present a model to quantify the information capacity of attention~\cite{hou2006time}. In this model, Hou and Zhang propose that visual attention dynamically adjusts its resolution over time and that this relationship can be described quantitatively. Their findings demonstrate a clear relationship between reaction time and the spatial resolution of attention~\cite{hou2006time}. As emphasized by Schewior and Lindner (2024), response time serves as an important indicator of cognitive processes in multimedia testing formats~\cite{schewior2024revisiting}. In the context of multimedia learning and testing~\cite{schewior2024revisiting, lindner2017identifying}, representational pictures (RPs) is a term that describes images that depict conceptually relevant content and aim to support understanding by visually representing the underlying subject matter~\cite{schewior2024revisiting}, in our interpretation: visual-graphical representations.

The review by Schewior \& Linder (2024) shows that, performance-based studies on the effects of learning with visual-graphical representations often show mixed results, while process-based indicators such as time on task can provide deeper insights into the cognitive impact of visual-graphical representations. Several studies found no significant changes in time on task when visual-graphical representations were included compared to text-only materials (e.g., \cite{schewior2024revisiting, ehrhart2023computer, lindner2017identifying, zheng2012solving}), while others reported either shorter (\cite{lindner2017merits, sass2012pictures}) or longer task durations~\cite{schewior2024revisiting, berends2009effect}. However, eye-tracking data reveal that learners spend less time processing textual elements and instead focus more on visual components when visual-graphical representations are present~\cite{schewior2024revisiting, lindner2017identifying, ogren2017there}. This shift in attention suggests that visual-graphical representations may not change overall task time but do influence how time is allocated during problem solving. Furthermore, visual-graphical representations appear to facilitate the construction of mental representations in early problem phases and aid in updating them during later stages~\cite{lindner2017identifying,schewior2024revisiting}. These findings highlight that even when time-on-task effects are ambiguous, visual-graphical representations can have measurable effects on learners’ cognitive processes.

From the perspective of Cognitive Load Theory~\cite{sweller1998cognitive}, tasks with high element interactivity require more working memory resources due to increased intrinsic cognitive load. According to Sweller et al. (1998, 2019), the interactivity of elements in a task determines the intrinsic cognitive load, as multiple interacting elements must be processed simultaneously and integrated meaningfully~\cite{sweller1998cognitive, sweller_cognitive_2019}. This suggests that tasks with high element interactivity are likely to require greater working memory resources, which may be reflected in increased processing time. Depending on the context of the task, longer processing times may be associated with increased cognitive effort, deeper engagement, or conversely with uncertainty and difficulties in understanding. As such, processing time should be interpreted with caution and always in relation to task performance and cognitive load. 
\\

\paragraph{Cognitive Load:} In addition to the processing time, the cognitive load can provide information on the processing of the information. According to Sweller et al. (2011), various methods exist to measure cognitive load, including performance measures, secondary tasks, physiological indicators, and subjective rating scales~\cite{Sweller2011}. The latter – such as the Cognitive Load Test developed by Klepsch et al. (2017) – assess cognitive load retrospectively and may be influenced by learners’ self-assessment and self-concept~\cite{Sweller2011, klepsch_development_2017}.
\\
\paragraph{Self-assessment:} In addition, learners' self-assessment (e.g. confidence in their answers) can serve as a metacognitive indicator of meaningful learning processes~\cite{schewior2024revisiting,lindner2021integrative}. Concepts that are processed with a higher level of subjective confidence are more likely to be stored in mid-term memory and can be retrieved. 
A comprehensive understanding of learning therefore requires not only outcome-oriented data but also process-oriented data, such as processing time or mental effort during the learning phase.

\subsection{B. Mechanisms/Aspects of qubit representations} 
\label{secsec:levesB1}

In order to enhance comprehension of the manner in which visual-graphical qubit representations can facilitate learning in \ac{qp}, a comprehensive category system was developed~\cite{qerimi2025exploring}. This system was employed in the structuring and evaluation of pivotal representational features in our previous research. The category system, which is grounded in the \ac{deft} framework (Ainsworth, 2006) and has been extended by findings from quantum education, was applied in an expert rating~\cite{ainsworth_deft_2006}. In this study, four representations (Bloch sphere, Quantum Bead~\cite{huber_beads_2024}, Circle Notation~\cite{johnston_programming_2019,bley_visualizing_2024}, and a pi chart model: Qake~\cite{donhauser2024}) were evaluated by experts across 16 features based on their perceived potential to facilitate learning.
According to the ratings, representations differ in how well they addressed central quantum concepts such as quantum state, superposition, and measurement~\cite{qerimi2025exploring}. In particular, the experts evaluated whether and how well the representations depict phase and amplitude -- two essential components of understanding QP. The Bloch sphere and Circle Notation were rated as more suitable for visualizing relative phase than the Quantum Bead and Qake representations.
Another category assessed by the experts was visual salience, which is the extent to which visual elements draw attention. Salience refers to how strongly a stimulus stands out in a particular context and attracts attention~\cite{higgins1996activation}. In educational contexts, this characteristic is crucial, as salient stimuli can effectively direct learners’ attention and support cognitive processing\cite{parr2019attention, rumbaugh2007salience, itti_computational_2001}. According to Cowan (1999), salient elements play a central role in the control of attention within working memory~\cite{cowan1999embedded}.

In this context, the Quantum Bead emerged as a particularly noteworthy element, garnering high ratings from experts who recognised its 3D design and utilisation of colour as key characteristics contributing to its prominence.

Overall, the expert evaluations make clear that different representations have different strengths and limitations, depending on the learning goal, the conceptual focus, and the visual and cognitive accessibility. It is evident that these insights provide a valuable basis for the selection or design of representations that are not only visually engaging but also educational effective. In order to facilitate a more profound comprehension of the manner in which the insights of the experts can be applied by learners, the following study was conducted. 

Based on differences identified through expert evaluations, specifically in \textit{Phase, Amplitude, Salience} and \textit{Concepts} on the representations: the Bloch sphere and the Quantum Bead, the study was conducted. To examine whether and how the expert-identified differences between the representations translate into differences in student learning, we used application-oriented tasks involving each representation. These tasks were analyzed with regard to processing time. We also measured content knowledge \cite{bitzenbauer_design_2024, waitzmann_wirkung_2023, hu2024investigating} and cognitive load~\cite{klepsch_development_2017} to learn more about the effects of different qubit representations on students' learning and the mid-term effects.

\section{III. Method} \label{sec:levelC}

\subsection{A. Piloting}  
A pilot study was carried out with N=8 participants (N=3 in group Bloch sphere and N=5 in group Quantum Bead) to ensure that the learning material and test items were neither too difficult or too easy. The necessity to modify the material and items was rendered obsolete by the students' feedback, which indicated that the material was comprehensible and the items were neither excessively challenging nor unduly simplistic. The methodology employed in the pilot study was replicated for the main study, as no adverse effects were observed in either case.

\subsection{B. Participants}
The sample in the main study comprised $N=149$ upper secondary school students in Bavaria. The participants were distributed across class level 11 ($n=46$), level 12 ($n=32$) and level 13 ($n=52$) (between the ages of 15 and 18). For $n=20$ participants, no information on the class level was available. Participants were primarily recruited through teachers who had scheduled visits to the student lab (PhotonLab) at the Max Planck Institute of Quantum Optics. These teachers were contacted in advance and asked whether they would be willing to participate in the study with their classes. Upon agreement, the entire class took part, provided that students presented a signed consent form from their legal guardians.

\subsection{C. Study design and process}
\begin{figure}[h!]
  \centering
  \includegraphics[width=0.5\textwidth]{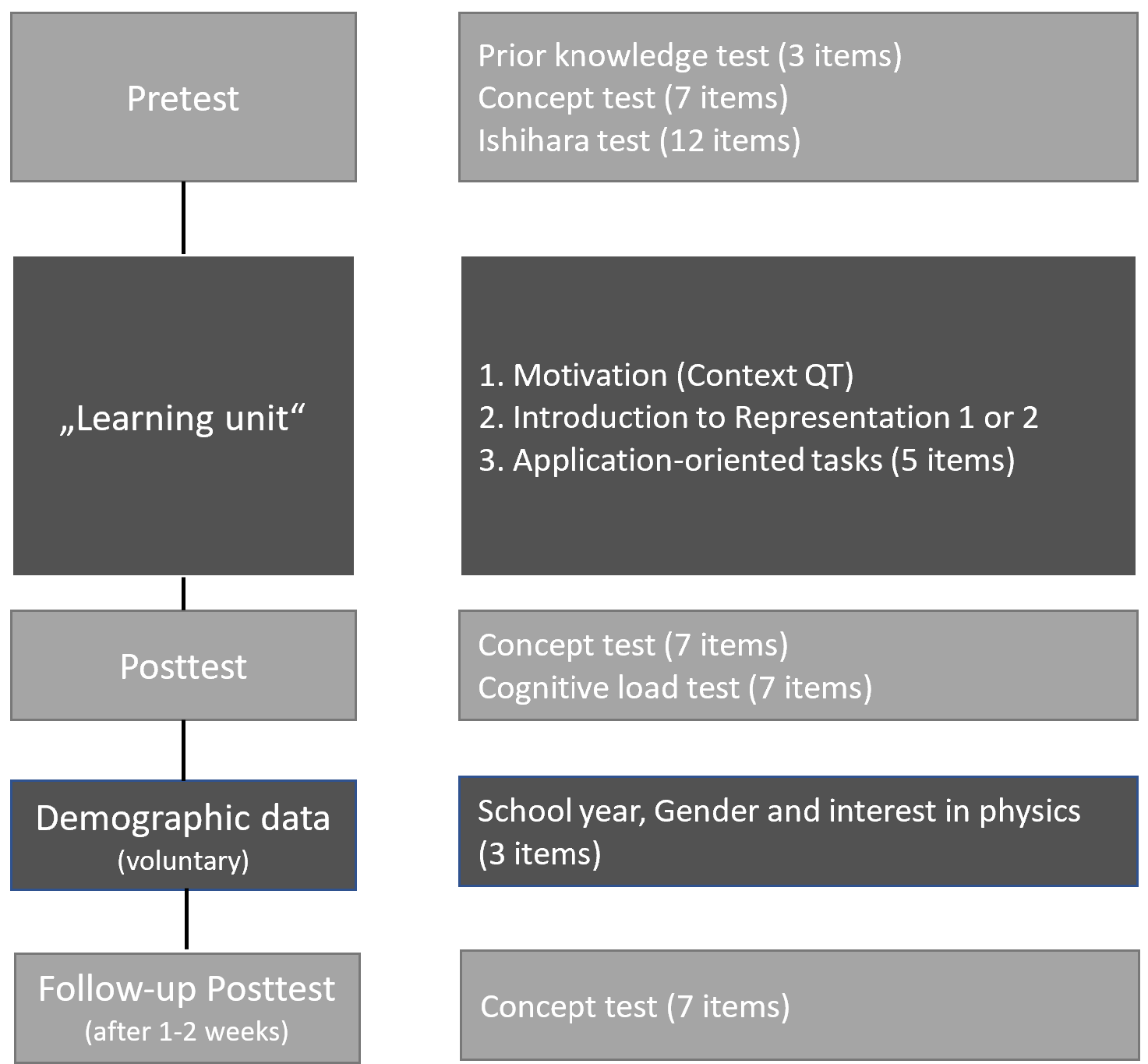} 
  \caption{Schematic overview of the process: After pseudonymised registration and a colour perception screening (Ishihara test), the participants completed a prior knowledge test and a conceptual pretest. This was followed by a learning unit that included an introduction to the respective representation and application-oriented tasks. The participants then completed a posttest and an assessment of cognitive load (based on Klepsch et al., 2017)~\cite{klepsch_development_2017}. A follow-up ($\text{Posttest}_2$) test was conducted 1–2 weeks later to assess medium-term retention performance.}
  \label{study_structure}
\end{figure}

The study followed a structured sequence (see Figure~\ref{study_structure}).

The study was designed as a mixed design with a between-subjects factor (representation: \textit{Quantum Bead} or \textit{Bloch sphere}) and a within-subjects factor (testing time: pretest, posttest, follow-up-test ($\text{Posttest}_2$)). 

First, all participants completed a pseudonymised survey. A colour perception test (Ishihara test) was conducted \footnote{\url{https://www.de.colorlitelens.com/farbenblindheit-test.html#TEST}} to rule out possible difficulties in interpreting the colour-coded Quantum Bead representations. If participants have three or more errors in the test, participants in the group of quantum beads were excluded~\cite{birch1997efficiency}. Following this, prior knowledge was assessed using three items related to waves, electromagnetic radiation, and beam splitters (items were used\cite{waitzmann_wirkung_2023}. In the pretest phase, students responded to conceptual questions on QP, which were later repeated in the posttest to evaluate learning gains and the follow-up test (items were used~\cite{bitzenbauer_design_2024, hu2024investigating, waitzmann_wirkung_2023}. A learning unit followed, consisting of a motivational introduction and an explanation of the assigned visual representation (Bloch sphere or Quantum Bead). After that, application-related tasks were carried out with the corresponding representations of the concepts, quantum state, superposition, quantum measurement, quantum measurement/quantum state, phase and amplitude. The time taken to complete the tasks was measured in milliseconds [ms]. After the learning phase, the posttest was conducted. To assess cognitive load during the learning process, items based on Klepsch et al. (2017) were administered~\cite{klepsch_development_2017}. Finally, a follow-up test ($\text{Posttest}_2$) was carried out 1–2 weeks later to examine the retention of knowledge over time.

\subsection{D. Selection of representations and Materials of the Learning unit}
\begin{figure}[htbp]
  \centering
  \includegraphics[width=\linewidth]{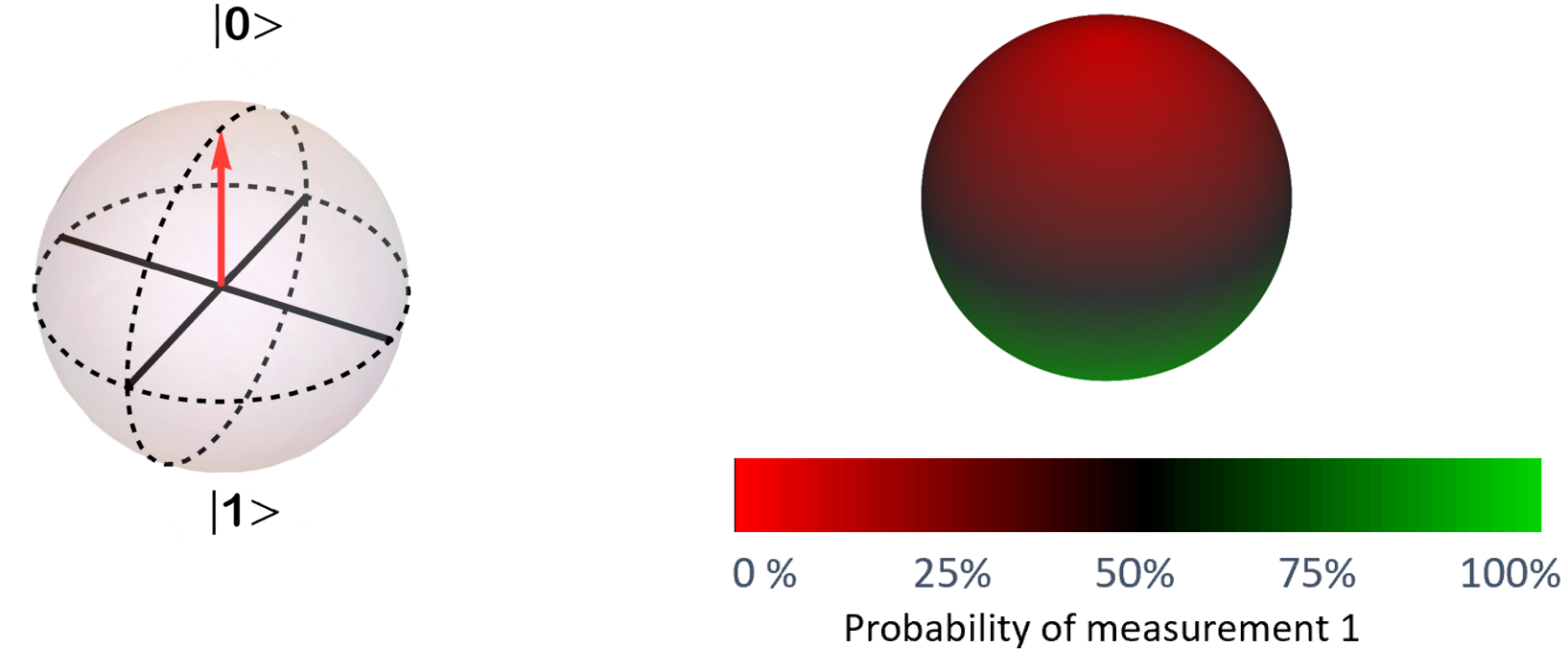}
  \caption{The Bloch sphere is shown on the left, while the Quantum Beads and their bar chart are on the right. The latter serves as a guide to the meaning of the colour changes. Images are self-created. Quantum Bead adapted from \cite{huber_beads_2024}}
  \label{Bloch_QB}
\end{figure}

To ensure a fair comparison, two three-dimensional representations were used that differed primarily in the specific features under investigation. The \textit{Quantum Beads} use colour-coded structures to visualise quantum concepts, while the \textit{Bloch sphere} uses the classic spherical representation with vector arrows. Concrete: The Bloch sphere represents the state of a single qubit as a vector on the surface of a unit sphere, offering a geometric interpretation of superposition and phase through spatial rotation. In contrast, Quantum Beads visualize quantum states using a red-green color gradient mapped onto a sphere: red indicates a high probability of measuring |0⟩, green corresponds to |1⟩ -- each relative to the conventional Z-axis measurement. Intermediate colors, such as black, represent a 50/50 superposition. As the quantum state evolves, the sphere rotates, and the visible color at the top changes accordingly (see Figure \ref{Bloch_QB}). 

\begin{figure}[htbp]
  \centering
  \includegraphics[width=\linewidth]{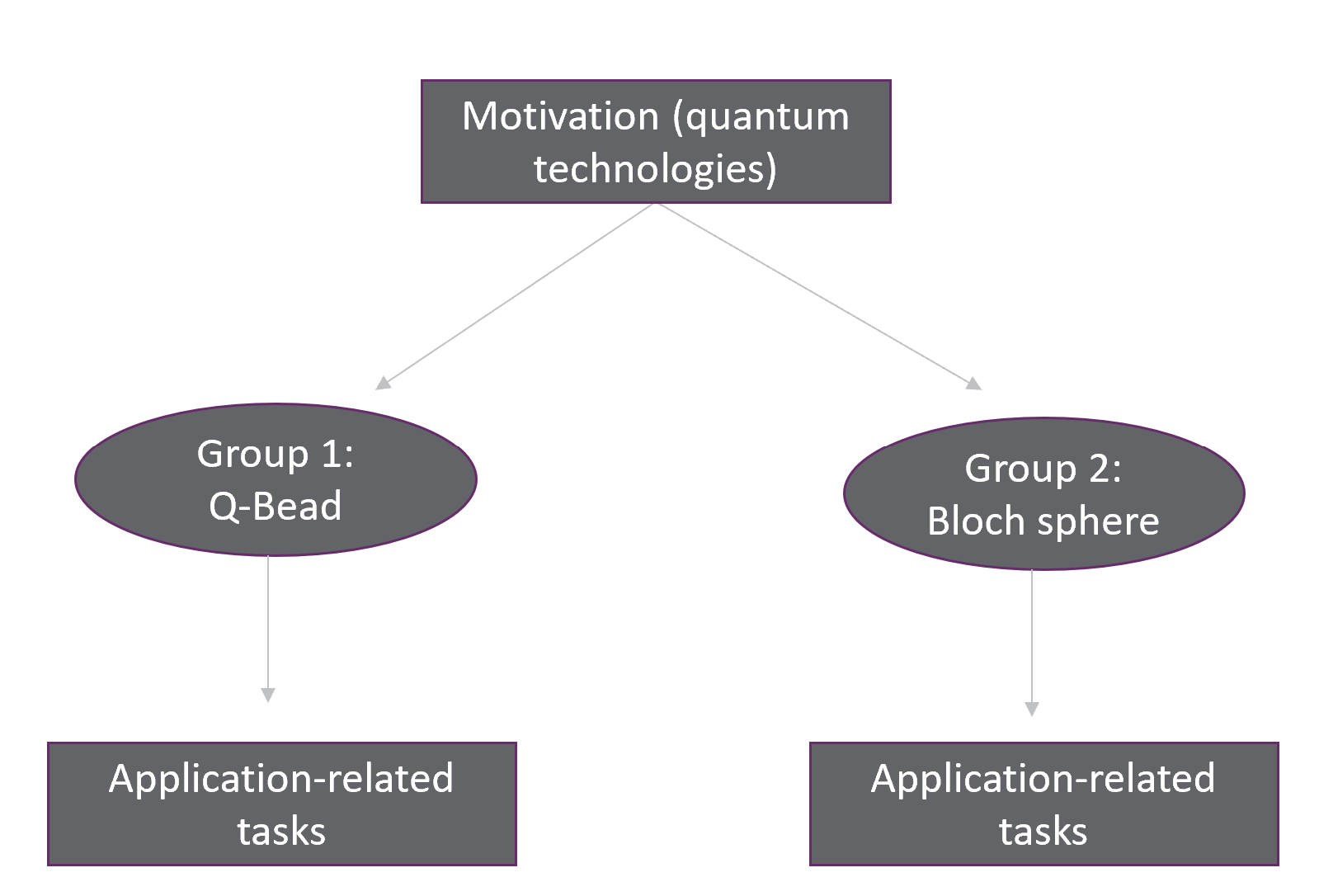}
  \caption{Structure of the learning unit and the application-oriented tasks.}
  \label{learning_unit}
\end{figure}

The learning materials were delivered digitally via iPads and were identical in content, structure and language, differing only in the way quantum information was represented. The unit was embedded in SoSci Survey (\hyperlink{SoSci Survey}{https://www.soscisurvey.de/en/index}).
Participants in group 1 received a version with the Quantum Bead, while group 2 worked with the Bloch sphere. Both groups start with the same context about quantum computing and the relevance of qubits, central concepts of QP, including quantum states, superposition, quantum measurement, phase (in form of a phase gate) and amplitude, followed by the introduction of the relevant representations (see Figure~\ref{learning_unit}). The explanations were kept simple and adapted to the prior knowledge of the secondary school students.

\begin{figure}[ht]
  \centering
  \includegraphics[width=0.45\textwidth]{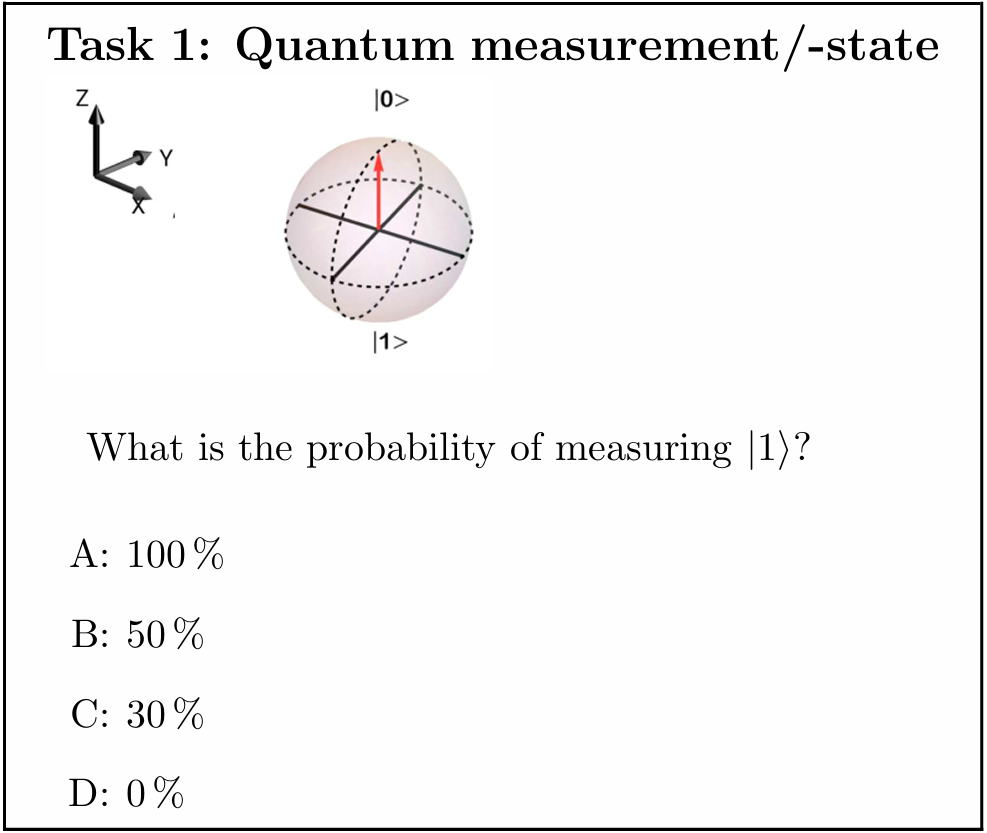}
  \caption{Task 1: Quantum measurement/ -state for the Bloch sphere group.The same task was set for Quantum Bead.}
  \label{fig:frage1}
\end{figure}

After the introduction, the participants worked on five application-oriented tasks directly related to the respective representation. The tasks could only be solved once the essential information had been extracted from the representation. The tasks were designed as multiple-choice questions and targeted specific aspects of the representations (e.g., spatial orientation, probability distribution, rotations, phase and amplitude). For each task, the processing time in milliseconds [ms] was recorded to allow for process-oriented analysis. An example task requiring participants to determine the probability of measuring $\ket{1}$ using the \textit{Bloch sphere} is shown in Figure~\ref{fig:frage1}.

The material for the learning unit, including subsequent application-oriented tasks (example in figure \ref{fig:frage1}, could be made available on request (Note:The material is in german language.

\subsection{E. Statistical Methods}

Various statistical methods were used to answer the research questions and test the hypotheses.
First, an \ac{anova} was planned to examine the influence of the different forms of presentation (Quantum Bead vs. block sphere) on learning success, while at the same time controlling for the effects of covariates such as prior knowledge or pre-test performance. However, due to the violation of the homogeneity of the regression slopes, a classic ANCOVA was not used and a multiple linear regression model was used instead. This allowed the predictors and their effects on the post-test score to be modelled separately. 

To evaluate the application-oriented tasks, we considered three variables: the number of correctly solved tasks (task accuracy), the total time spent on all tasks in seconds (task time), and the ratio of correct solutions to time (efficiency). Since the data did not meet the assumptions for an ANOVA—particularly due to a lack of normal distribution—we used Mann–Whitney U tests to compare the groups.

A multivariate analysis of variance (MANOVA) was conducted to the effects of representation type on cognitive load. Three dependent variables were considered simultaneously: \textit{intrinsic cognitive load (ICL)}, \textit{extraneous cognitive load (ECL)} and \textit{germane cognitive load (GCL)}. This procedure is particularly suitable for investigating multivariate effects when it is suspected that the dependent variables interact with each other, as is the case with the cognitive load dimensions. 
In the event of significant main effects, subsequent post-hoc tests were carried out, whereby the Bonferroni correction was applied to control for the alpha error in multiple comparisons.

The effect sizes of the results were given in Cohen's $d$ to better classify their practical relevance. This measure allows the size of an effect to be estimated independently of the sample size. In case of violations of the normal distribution assumption, in particular with smaller samples or skewed distributions, Wilcoxon tests were used as a non-parametric alternative to the t-test in order to be able to make robust statements about differences between two measurement points or groups.

To evaluate whether group differences emerged in medium-term retention, an Analysis of Covariance (ANCOVA) was conducted. In this model, the dependent variable was the test performance in the follow-up posttest (Posttest$_2$). The independent variable was the representation group (Quantum Bead vs. Bloch sphere), and the covariate was the performance in the immediate posttest (Posttest$_1$), which served to control for individual differences in initial learning successes.

To analyse the development of learning over time, not only between Posttest$_1$ and Posttest$_2$, but across the entire learning process, a mixed ANOVA with repeated measures was conducted. The within-subjects factor was time (Pretest, Posttest$_1$, Posttest$_2$), and the between-subjects factor was the representation group (Quantum Bead vs. Bloch sphere).
In addition, a separate mixed ANOVA with repeated measures was performed for response confidence across the same three time points, in order to investigate changes in learners’ perceived confidence over the course of the study. 

All statistical analyses were performed in R (version 4.4.0; R Core Team, 2024), and the corresponding analysis scripts are available upon request.

\section{IV. Results} \label{sec:levelD}

\subsection{A. Sample considered}

The final sample consisted of $N = 146$ students. Due to suspected red-green color vision deficiency, three participants were excluded from the analysis~\cite{birch1997efficiency}. Of the remaining participants, 85 identified as male, 46 as female, 1 as diverse, and 14 did not provide gender information. Group allocation was balanced: 75 students were randomly assigned to the Quantum Bead group (23 female, 46 male, 1 diverse, 5 no gender specified), and 71 to the Bloch sphere group (23 female, 39 male, 0 diverse, 9 no gender specified).

\subsection{B. RQ1 – Learning Outcome Based on Representation Type}\label{sec:results_rq1}

Pre-Post Comparison: Wilcoxon signed-rank tests revealed significant learning gains in both groups from pretest to posttest in Bloch sphere ($V= 238 $, $p < .001$) and in the Quantum Bead ($V= 370 $, $p < .01$) group, indicating that students improved regardless of representation.

To investigate whether learners’ outcomes differ as a function of the visual representation used (Quantum Bead vs. Bloch sphere), we first explored pre-post differences and then conducted a multiple linear regression analysis controlling for covariates.

\paragraph{Multiple Linear Regression (MLR).}
To test the effect of the representation while controlling for prior knowledge and pretest performance, we conducted a multiple regression analysis with the posttest score as dependent variable. The model included the predictor variables \textit{group}, \textit{pretest}, and \textit{prior knowledge}. The regression model was statistically significant overall, $F(3, 142) = 29.21$, $p < .001$, explaining $R^2 = .38$ of the variance in posttest scores (adjusted $R^2 = .37$).

\begin{equation}
Y_i = 0.42 + 0.08 \cdot X_{1i} + 0.55 \cdot X_{2i} + 0.30 \cdot X_{3i} + \varepsilon
\end{equation}

\noindent
{\footnotesize
 $Y_i$ = Posttest score; $X_{1i}$ = Group (0 = Bloch, 1 = Q-Bead); $X_{2i}$ = Pretest score; $X_{3i}$ = Prior knowledge; $\varepsilon$ = residual error.}

\vspace{0.5cm}

As shown in Table~\ref{tab:mlr_coefficients}, pretest performance ($p < .001$) and prior knowledge ($p = .008$) were significant predictors of posttest outcomes. The representation group, however, had no significant effect ($p = .703$). Thus, participants with higher pretest and prior knowledge scores tended to perform better on the posttest, regardless of the representation used.

\begin{table}[h!]
\centering
\caption{Multiple Linear Regression Coefficients Predicting Posttest Score}
\label{tab:mlr_coefficients}
\begin{tabular}{lcccc}
\toprule
Predictor & Estimate & Std. Error & $t$ & $p$ \\
\midrule
(Intercept) & 0.420 & 0.526 & 0.80 & .426 \\
Group ($X_1$)$^*$ & 0.083 & 0.217 & 0.38 & .703 \\
Pretest Score ($X_2$) & 0.547 & 0.068 & 8.00 &  *** \\
Prior Knowledge ($X_3$) & 0.300 & 0.112 & 2.68 & ${**}$ \\
\bottomrule
\end{tabular}

\vspace{0.5em}
\raggedright
{\footnotesize
$^*$ Dummy coding was introduced to correctly account for the group variable (categorical variable) in the regression equation.\\
Significance levels: \textbf{***} $p < .001$, \textbf{**} $p < .01$, \textbf{*} $p < .05$}
\end{table}

\paragraph{Summary:}
Overall, no direct effect of representation on posttest performance was observed. Learning gains occurred in both groups, and individual differences (pretest, prior knowledge) had stronger predictive value than the type of representation.

\subsection{C. RQ1 - Task Performance} \label{sec:task_performance}

To evaluate the application-oriented performance when learning with the visual-graphical representations, participants solved five applications-oriented tasks addressing key quantum concepts: quantum state, superposition, quantum measurement, as well as phase gate and amplitude with the representations (see Figure \ref{study_structure}). Five tasks were set for this. The fourth, the phase gate task was excluded. The task did not differentiate between groups, indicating that it may not have been sensitive enough to capture group-specific effects of the representation. It was excluded for further analysis. 

For the remaining tasks, the following indicators were calculated for each participant:

\begin{itemize}
    \item \textbf{Task accuracy:} Number of correctly solved tasks (0--4)
    \item \textbf{Task time:} Total time spent on all tasks (transferred in seconds [s])
    \item \textbf{Efficiency:} Correct solutions divided by time
\end{itemize}

\begin{figure}[htbp]
    \centering
    \includegraphics[width=0.5\textwidth]{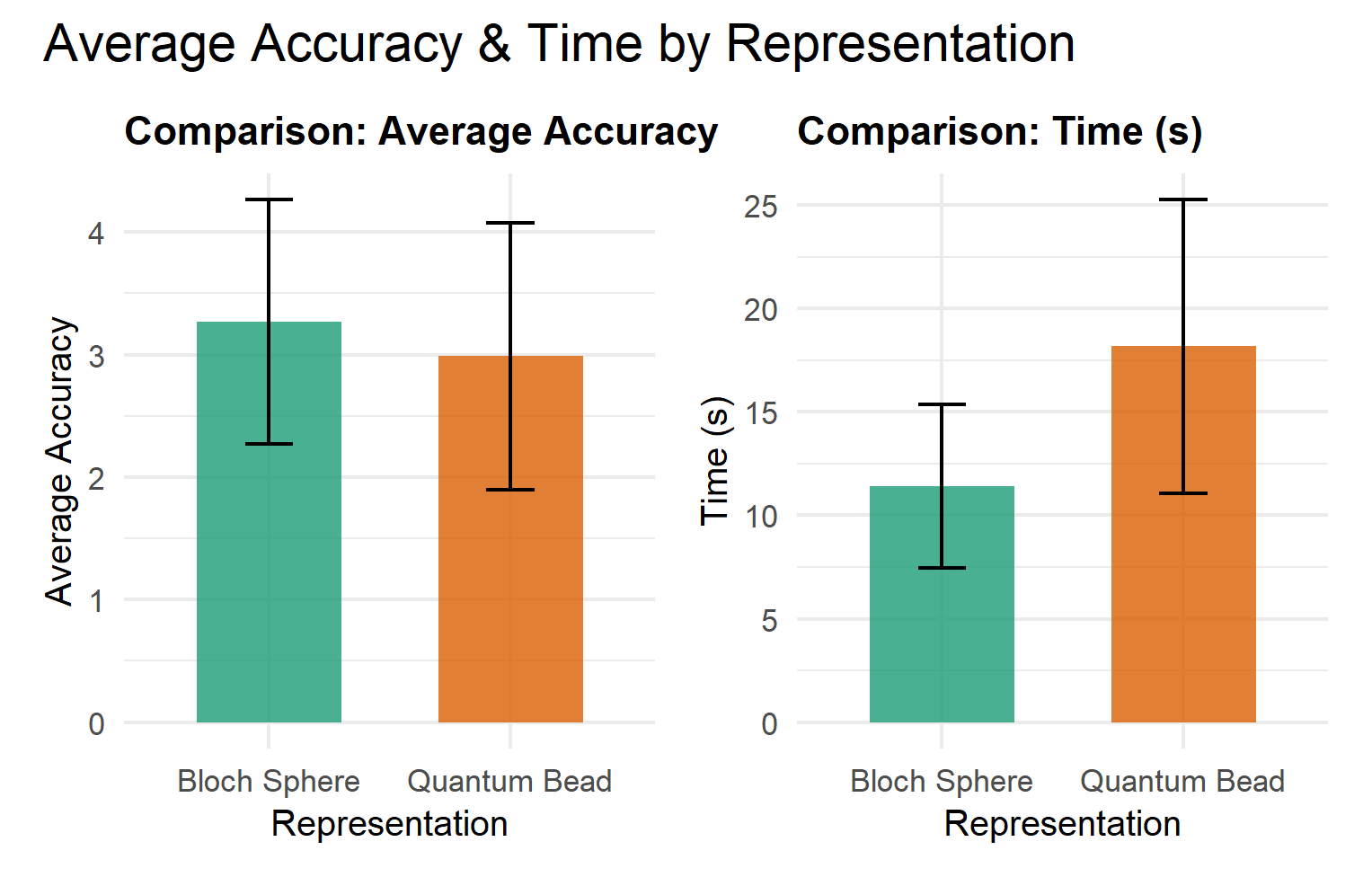}
    \caption{Mean response time (seconds) and the \ac{sd} after representation. The Bloch sphere group showed significantly shorter processing times.}
    \label{fig:antwort_zeit}
\end{figure}

\paragraph{Task accuracy and Task time:} Descriptive results showed that participants in the Quantum Bead group solved on average {[2.99]} tasks correctly ({SD = [1.09]}), compared to {[3.27]} in the Bloch sphere group ({SD = [1.00]}). The average task time was {[11.40] s} in the Bloch sphere group ({SD = [3.95]}), and {[18.20] s} in the Quantum Bead group ({SD = [7.10]}). No significant difference was found between the two representations for response accuracy ($W = 2906.5$, $p = .090$).
However, a highly significant difference was found for the response time ($W = 882$, $p < .001$, $r = .56$), with the group with the Bloch sphere completing the tasks significantly faster. This is also clearly shown in Figure ~\ref{fig:antwort_zeit}.

\paragraph{Efficiency:}
The efficiency (correct answers per second) was slightly higher in the Bloch sphere group (\textit{M = [0.323], SD = [0.162]}) than in the Quantum Bead group (\textit{M = [0.187], SD = [0.099]}). The Mann–Whitney U test revealed a significant difference in efficiency in task performance between the groups (W = 3642, $p < .001$), with a medium effect size according to Cohen (r = .48).

\begin{figure}[htbp]
    \centering
    \includegraphics[width=0.4\textwidth]{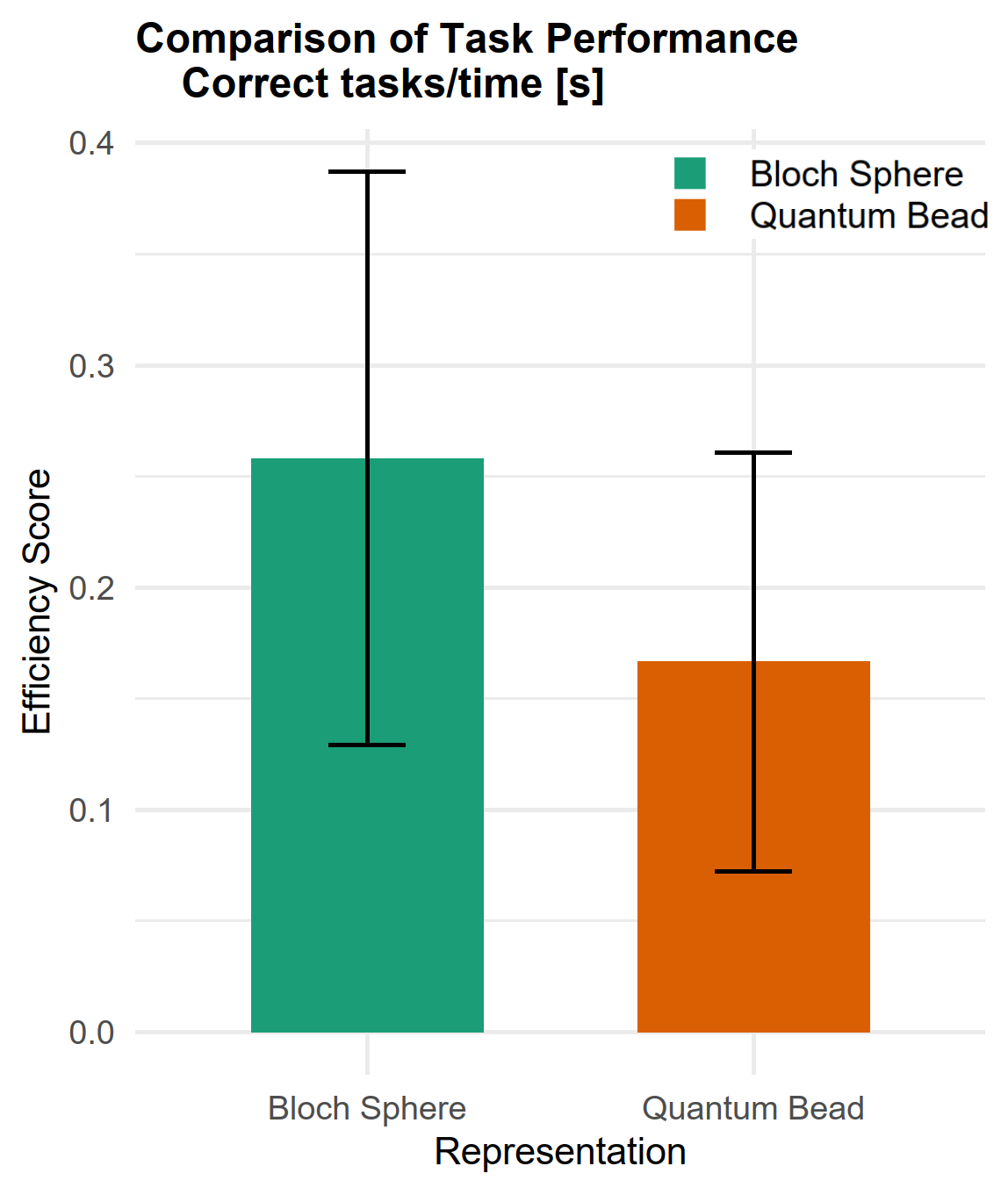}
    \caption{Mean efficency score and SD: The tasks were answered much more efficiently by pupils as soon as they were presented with the Bloch sphere. The efficiency score is significantly higher in the Bloch sphere group.}
    \label{fig:eff}
\end{figure}

\paragraph{Summary:}
Task accuracy and task time were also analyzed separately. While no significant difference was found between the two groups in terms of accuracy, a highly significant difference was observed in task time: participants in the Bloch sphere group completed the tasks substantially faster. Consequently, learning processes with the Bloch sphere representation were more efficient—measured as correct answers per second. 

\subsection{D. RQ2: Cognitive Load differences between groups}

To investigate whether different visual-graphical representations lead to a more effective use of cognitive resources, a multivariate analysis of variance (MANOVA) was conducted. The independent variable was study group (Quantum Bead vs. Bloch sphere), and the dependent variables were three types of cognitive load: intrinsic cognitive load (ICL), extraneous cognitive load (ECL), and germane cognitive load (GCL).

\begin{figure}[htbp]
    \centering
    \includegraphics[width=0.4\textwidth]{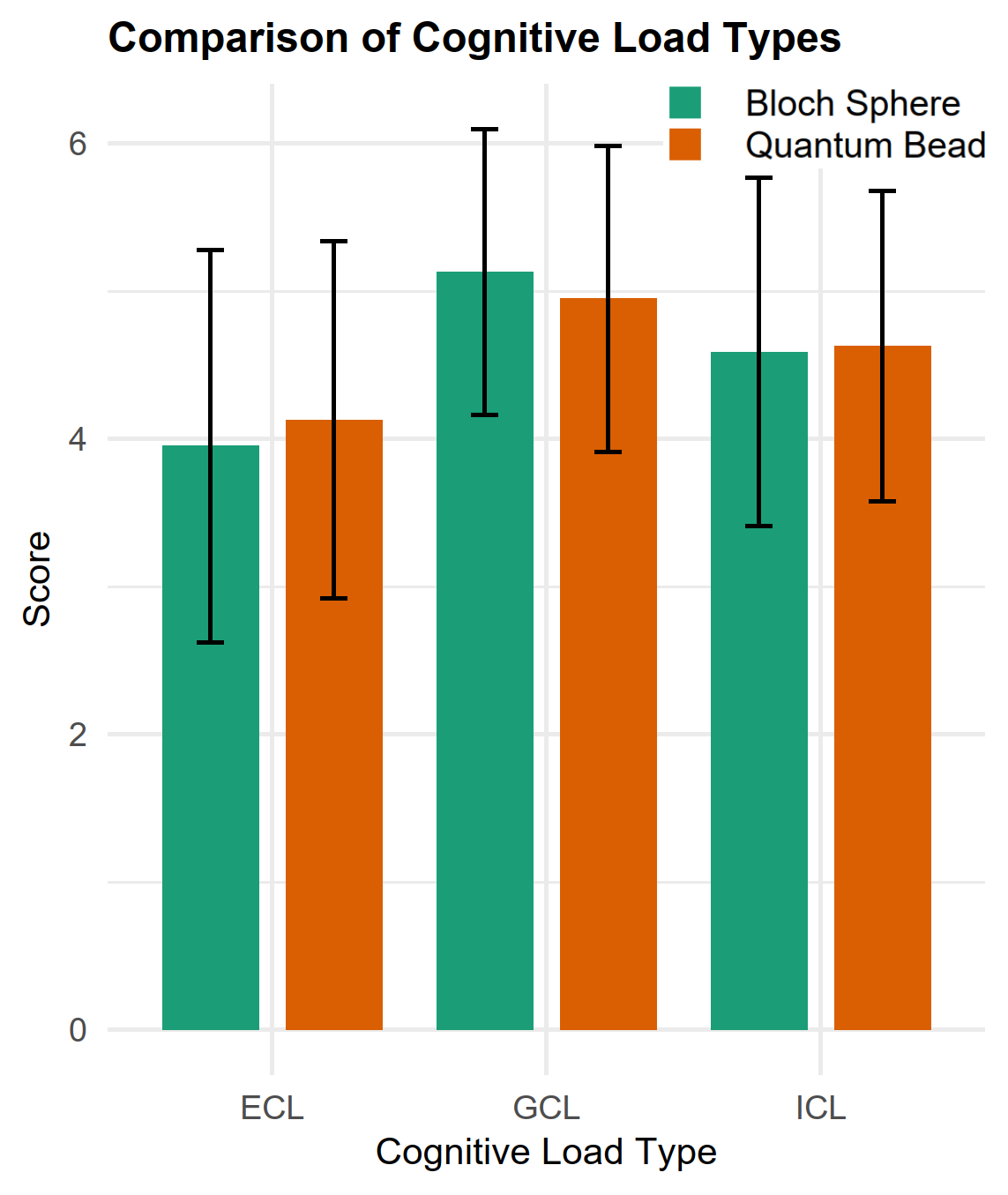}
    \caption{Mean cognitive load score and SD. There were no significant differences between the groups.}
    \label{fig:cognitive}
\end{figure}

Descriptive statistics indicated only slight differences between the groups. For ICL, the Bloch sphere group had a mean of $M = 4.59$ ($SD = 1.18$), while the Quantum Bead group had a mean of $M = 4.63$ ($SD = 1.04$). Regarding ECL, participants in the Bloch sphere group reported $M = 3.95$ ($SD = 1.33$), and the Quantum Bead group $M = 4.17$ ($SD = 1.25$). In terms of GCL, the Bloch sphere group scored slightly higher ($M = 5.13$, $SD = 0.97$) than the Quantum Bead group ($M = 4.94$, $SD = 1.03$).

The results of the MANOVA revealed no statistically significant overall effect of group on the combined cognitive load measures, $Pillai's~Trace = 0.0186$, $F(3, 148) = 0.90$, $p = .440$.

\paragraph{Summary:} 
\begin{itemize}
\item Hence, no significant group differences were found for any type of cognitive load. 
\item These findings suggest that the use of different visual-graphical representations did not lead to significantly different cognitive load experiences between the groups.
\end{itemize}

\subsection{E. Results for RQ3: Medium-Term Retention of Quantum Concepts}

To examine potential differences in the retention of information, participants were asked to complete the same posttest items again approximately 1–2 weeks later. Using their pseudonymised codes, responses could be matched to the initial data set. As not all participants completed the follow-up test, the sample size was accordingly reduced. 

To assess medium-term retention, an ANCOVA was conducted using the follow-up posttest score ($\text{Posttest}_2$) as the dependent variable. The independent variable was the representation group (Quantum Bead vs. Bloch sphere), with the immediate posttest score ($\text{Posttest}_1$) as covariate.

\paragraph{Sample:} The analysis included $N=98$ students (Group 1: $n=53$, Group 2: $n=45$). Group 1 included 15 females, 32 males, 1 diverse, and 5 participants without gender information. Group 2 consisted of 17 females, 20 males, and 8 without gender information.

The covariate $\text{Posttest}_1$ significantly predicted the follow-up posttest score ($\text{Posttest}_2$) ($F(1,97) = 18.82$, $p < .001$), indicating that higher scores in the immediate posttest were associated with better mid-term performance. However, there was no significant group effect on follow-up performance when controlling for prior performance ($F(1,97) = 0.03$, $p = .863$). The means were nearly identical between groups (Quantum Bead: $M = 3.44$, Bloch sphere: $M = 3.49$).

\begin{table}[h!]
\centering
\caption{ANCOVA results for predicting mid-term retention (Posttest\_2)}
\label{tab:ancova_retention}
\begin{tabular}{lcccc}
\toprule
Predictor   & Sum Sq & Mean Sq & $F$   & $p$       \\
\midrule
Posttest\_1 & 43.34  & 43.34   & 18.82 & ***       \\
Group       & 0.07   & 0.07    & 0.03  & n.s.      \\
Residuals   & 223.43 & 2.30    &       &           \\
\bottomrule
\end{tabular}

\vspace{1mm}

\small
\textit{Note.} *** $p < .001$; n.s. = not significant. The covariate Posttest\_1 significantly predicted Posttest\_2 performance. No group differences were observed after controlling for Posttest\_1.
\end{table}

\subsubsection{Mixed ANOVA with repeated measure: Development of learning process}
To examine the development of conceptual understanding over time, a mixed ANOVA with repeated measure was conducted with \textbf{time} as a within-subjects factor \textit{(Pretest~$\rightarrow$~Posttest$_1$~$\rightarrow$~Posttest$_2$)}.
 
The analysis yielded a significant main effect of time, $F(2, 188) = 4.23$, $p = .016$, $\eta^2_G = .013$, indicating a learning effect. No significant group difference was found, $F(1, 94) = 0.82$, $p = .818$, and the interaction time × group was also non-significant, $F(2, 188) = 0.10$, $p = .902$. Post-hoc analyses revealed significant differences between the pretest and both posttests ($p < .001$), but no difference between Posttest$_1$ and Posttest$_2$. The learning gains remained stable over time following the learning unit.

\subsubsection{Mixed ANOVA with repeated measure: Development of Confidence}

\begin{figure}[htbp]
    \centering
    \includegraphics[width=0.4\textwidth]{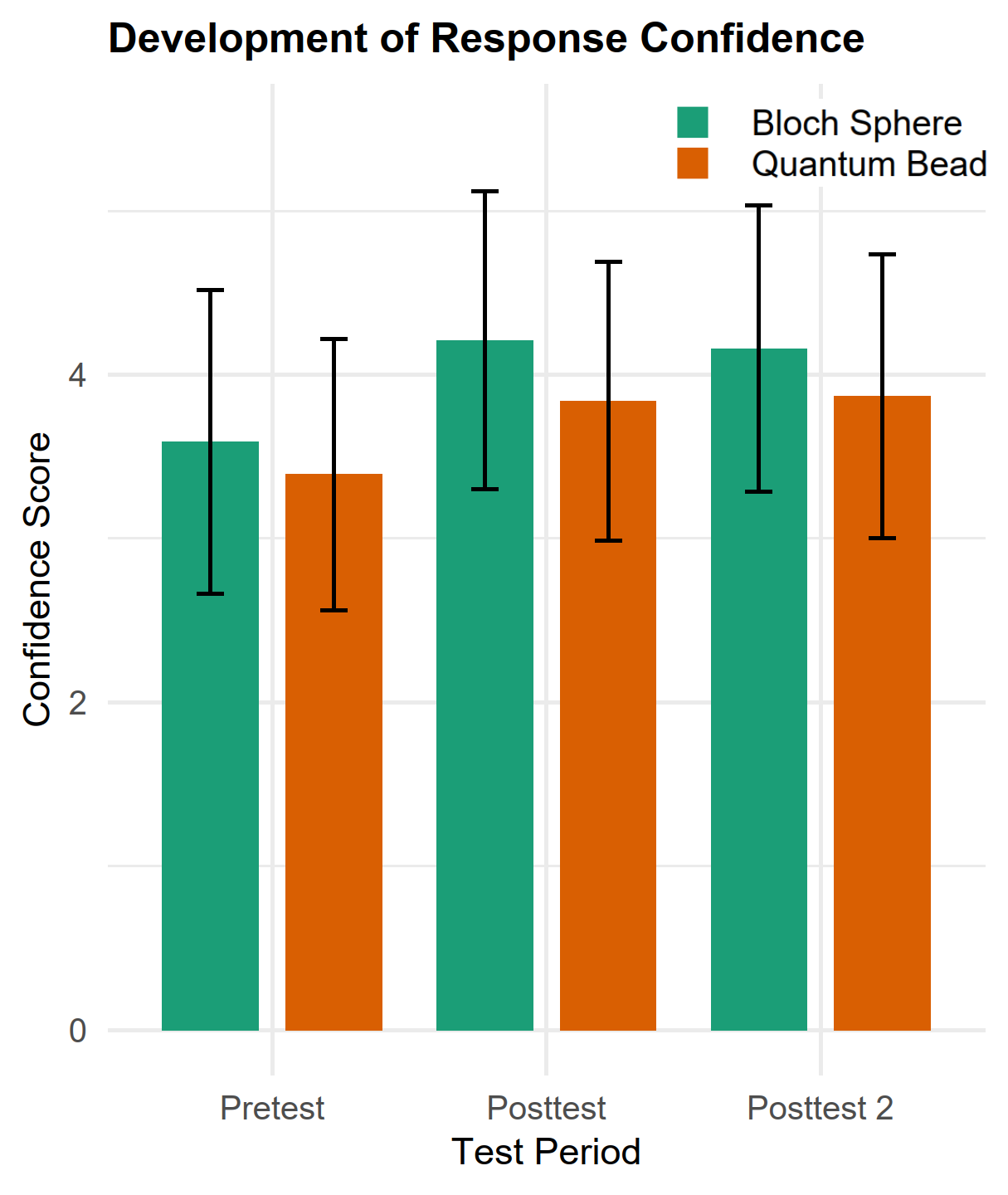}
    \caption{Response confidence over time and the SD. Confidence increased significantly, with no significant group difference after the learning unit.}
    \label{fig:antwortsicherheit}
\end{figure}

We also analyzed response security over time using a mixed ANOVA with repeated measures. It showed a significant main effect of the test time, $F(2,186) = 39.36$, $p < .001$,$ \eta² = .29$, indicating a significant increase in response certainty over time. Figure~\ref{fig:antwortsicherheit} shows a descriptive difference in response confidence between the groups after the learning unit; however, this difference is not statistically significant ($p = .066$).
Post hoc analyses revealed significant differences between pretest and both posttests ($p < .001$), but no difference between posttest and posttest 2 see Figure~\ref{fig:antwortsicherheit}. Confidence gains remained stable over time after the learning unit.

\paragraph{Summary}

\begin{itemize}
    \item \textbf{ANCOVA:} After controlling for Posttest\_1, no significant group difference in Posttest\_2 ($p = .863$); Posttest\_1 was a strong predictor of retention.
    \item \textbf{Learning process:} Significant improvement over time ($p = .016$); no difference between groups ($p = .821$).
    \item \textbf{Confidence:} Increased significantly after instruction ($p < .001$); remained stable over time; no group × time interaction.
\end{itemize}

\section{V. Discussion\label{sec:levelE}}

\subsection{A. Learning Outcome and Task performance}

A comparison of pre- and posttest scores revealed learning gains overall, but no significant differences between the two instructional groups. As both groups received identical textual explanations—differing only in the type of visual representation introduced—this suggests that while the inclusion of a representation may have supported learning in general, the specific representational features under investigation contributed only marginally to differences in conceptual learning outcomes.

However, process-based data from the application-oriented tasks on amplitude, superposition, quantum measurement, and quantum state revealed clear differences in processing time and efficiency. This indicates that learning with visual-graphical representations can indeed vary -- especially in terms of how learners process and apply information~\cite{sweller_cognitive_2019, schewior2024revisiting, lindner2017identifying}.

The results of our study support the assumptions made by experts in the prior expert rating: the Bloch sphere appears to be more effective for application-oriented \textit{tasks} involving representations. This is also reflected in the task results — learners in this group were significantly more efficient (task/time) in answering questions related to quantum state, amplitude, superposition, and quantum measurement.  

The results also suggest that the Bloch sphere enabled more efficient task processing, while the higher salience attributed to the Quantum Bead -- according to the expert ratings -- did not produce a compensatory effect.  Based on the qualitative data from the expert rating (free text responses), it was already noted that the category of salience poses a challenge in the rating process, as the necessary conditions for salience are difficult to achieve in a static rating context~\cite{qerimi2025exploring}. This issue is also reflected in the present study, which found that visual salience alone did not result better. Another possible explanation is that specific aspects such as phase difference or amplitude are more easily localized via the vector arrow in the Bloch sphere and helps to highlighted the relevant aspect better~\cite{sweller_cognitive_2019, mayer_signaling_2014}. This may allow learners to extract relevant information more quickly and efficiently compared to the Quantum Bead representation, where such features are less visually explicit. The clearer directional cues in the Bloch sphere may have helped to guide learners' attention more effectively toward conceptually relevant elements. 

The effectiveness of the Bloch sphere as a cognitive support medium for learning about quantum computing has already been highlighted in previous studies~\cite{hu2024investigating}. The more efficient processing of application-oriented tasks with the Bloch sphere may be attributed to the clearer and more structured visualization of key information--especially phase. The distinct direction vector and visible coordinate axes allow learners to extract relevant information more directly, for example when estimating measurement probabilities along the Z-axis. In such cases, a simple projection of the state vector onto the Z-axis is sufficient to infer the likely measurement outcome. 

In contrast, while the Quantum Beads were perceived as visually salient, their use of color gradients and state-dependent rotation appears to spatially delocalize relevant information. For instance, learners must interpret the color at the top of the bead—ranging from red to green—to estimate probabilities, which may be less immediate and require additional cognitive steps. This may increase cognitive load and lead to longer task processing times, as learners must first interpret and integrate the visual information.

On the other hand, studies suggest that increased attention can be associated with more intensive information processing (e.g. \cite{reynolds_influence_1982, mayer_fiorella_2014, cowan1999embedded}). Interestingly, however, the longer processing times in the Quantum Bead group did not lead to significant differences in pre-to-posttest learning gains. This suggests that while Quantum Beads may influence attention and processing time, they do not necessarily result in improved learning outcomes. One possible explanation is that the increased salience required learners to invest more cognitive resources, but without yielding additional learning benefits.

According to Cognitive Load Theory, the split-attention effect is particularly relevant in the context of text–image combinations~\cite{sweller_cognitive_1994}. This effect states that spatially separating related visual information sources -- such as text and image -- can increase cognitive load and thereby impair learning performance~\cite{sweller_cognitive_1994}. In this case, the students in the Quantum Bead group had to interpret both representations (bar and sphere) in order to arrive at a clear solution (see figure~\ref{Bloch_QB}). The reduced efficiency observed in the application-oriented task performance may thus be attributed to the split-attention effect. To mitigate this effect, the representations were placed in close spatial proximity. However, no differences in cognitive load were found.

These findings suggest that the instructional impact of visual representations should not only be assessed by learning outcomes, but also by their role in supporting cognitive processing and structuring task performance.

\subsection{B. Cognitive Load differences between groups}
This study investigated whether different visual-graphical representations (Bloch sphere and Quantum Bead) facilitate different effective use of cognitive resources when learning QP concepts.  Based on expert rating and cognitive load theory, we expected differences in cognitive load in particular~\cite{qerimi2025exploring, sweller_cognitive_1988, sweller_cognitive_1994}.
However, the MANOVA showed no significant group differences in intrinsic, extraneous or germane cognitive load. Although the Bloch sphere group reported slightly higher germane load and slightly lower extraneous load than the Quantum Bead group, these differences were not statistically significant. Therefore, the hypothesis was not supported by the data.

Although the efficiency results suggest that participants in the Bloch Sphere group processed information more quickly, this did not result in a significantly lower self-reported cognitive load. According to Sweller et al. (2019), differences in cognitive load should become apparent when information is easier to locate or process~\cite{sweller_cognitive_2019}. The absence of such differences in our data may be due to the limitations of the subjective rating method employed. Retrospective self-assessments, such as the one employed here, are sensitive to individual interpretation and prior expectations~\cite{Sweller2011}. In future studies, it may be advisable to use more objective measures of cognitive load, such as dual-task paradigms or eye tracking, or to use rating scales that are more closely tailored to the specific demands of the learning material. Alternatively, a within-subjects design, in which each participant works with multiple representations and compares their perceived cognitive load across different conditions, could offer greater sensitivity to subtle differences in cognitive demands.
An other possible explanation is that both representations were presented in a clearly structured and well-supported learning environment, which is likely to have reduced extraneous load in both groups. In addition, both representations were unfamiliar to the learners, which may have led to similar levels of cognitive processing in both conditions. In contrast to the expectations from the expert rating, the visual salience of the Quantum Bead alone did not lead to an efficient use of cognitive resources or enhanced processing—possibly because the learners lacked prior experience with this type of representation~\cite{cowan1999embedded, qerimi2025exploring}.
These findings suggest that the impact of a representation depends not only on its design, but also on how well it relates to learners' prior knowledge and how it is embedded in the instructional context. While previous studies (e.g., \cite{cowan1999embedded}; \cite{mayer2002multimedia}) have highlighted the role of salience in directing attention, this effect did not translate into measurable differences in cognitive load in our study.

\subsection{C. Medium-Term Retention of learning}

The analysis of medium-term retention (RQ3) revealed no significant differences between the representation groups after controlling for immediate post-test performance. This finding indicates that, while the representations varied in terms of processing efficiency during the tasks, there were no significant differences in the retention of the concepts over time.
The strongest predictor of follow-up ($\text{Posttest}_2$) performance was the immediate posttest result, suggesting that early learning success played a crucial role in mid-term retention. These results are consistent with long-established findings by Hattie, who showed that prior knowledge is a strong indicator of overall learning outcomes~\cite{hattie2013visible}.

The repeated-measures ANOVA showed that learning gains were maintained after the learning unit but did not increase further, pointing to a stabilisation effect rather than continued learning. Both groups improved over time and followed a similar learning trajectory, with no evidence that one group retained information more effectively than the other. With regard to learners' response confidence, a significant increase was observed over time, which remained stable after the instructional phase. Learners thus became more confident following the intervention, and this confidence was sustained in the follow-up. Although the Bloch sphere group showed slightly higher confidence values, the difference was not statistically significant ($F(1, 93) = 3.45$, $p = .063$), and both groups developed similarly. 

As noted in the review by Schewior\&Lindner (2024), the present findings similarly indicate that response confidence increased through the use of visual-graphical representations—regardless of group—and remained stable over time~\cite{schewior2024revisiting}.

\section{VI. Limitations}

The conclusions of the study are primarily be drawn based on process-related (time reaction) data regarding the processing time of the two representations (Bloch sphere vs. Quantum Beads). More detailed insights into the processing mechanisms -- such as fixation duration, gaze patterns, or visual attention distribution -- would have required the use of eye-tracking data. Using specific gaze data, it would have been possible to recognise how often and for how long the learners looked at the presentation until they answered the task.

It should also be noted that the majority of participants came from a school student group that had already shown an interest in QP prior to the study. Recruitment took place mainly in the context of scheduled visits to the student lab at the Max Planck Institute of Quantum Optics, which may have led to a certain degree of prior content exposure or motivation.

Another limitation concerns the short time (1-2 weeks) interval between the two posttests. It is possible that participants remembered their previous answers and responded accordingly, which may have influenced the measured stability of learning outcomes.

Furthermore, it is worth mentioning that working with the Q-Beads may have posed additional demands on learners, as they needed to integrate two representational elements (color bars and the Quantum Bead). Alternatively, some learners may have already internalised which color corresponds to which state—without this prior knowledge being explicitly controlled for in the study.

The evaluation of salience was based on an expert rating and did not reflect the subjective perception of the learners themselves. It therefore remains unclear whether visually highlighted elements were actually perceived and processed more strongly during learning.

\section{VII. Future Research}

Given that expert ratings revealed both strengths and limitations across all representations, future studies could build on this work by systematically comparing additional representational features~\cite{qerimi2025exploring}. For example, it would be valuable to examine which specific quantum concepts or instructional goals benefit most from the use of Quantum Beads in classroom settings. Furthermore, future research could explore learning scenarios in which multiple qubit representations are provided simultaneously, to investigate potential advantages of representational flexibility. To examine the impact of design features such as salience or visual complexity more directly, methodologies like eye tracking could be employed to gain deeper insight into learners' visual attention and processing strategies.

\section{VIII. Conclusion\label{sec:levelF}}

This study investigated the influence of two visual-graphical representations—the Bloch sphere and Quantum Beads—on students’ learning of QP concepts in the context of quantum computing. Overall, no significant differences in learning outcomes were observed between the two groups in the pre-post comparison.
However, clear differences emerged in process-related measures during task completion: participants in the Bloch sphere group completed application-oriented tasks significantly faster and more efficiently. This suggests that the Bloch sphere may better support cognitive processing, particularly in terms of quickly identifying and applying relevant information~\cite{sweller_cognitive_2019}. The more efficient processing of application-oriented tasks when using the Bloch sphere may be due to the fact that the distinct vector arrow and the visible coordinate axes within the sphere~\cite{sweller_cognitive_2019} – an advantage also confirmed by expert ratings in ~\cite{qerimi2025exploring}. In contrast, the higher visual salience of the Quantum Beads did not lead to improved learning outcomes or efficiency, but was associated with longer processing times—possibly due to increased cognitive demands when interpreting the representation. Rather, it can be assumed that the use of color gradients and sphere rotation delocalizes relevant information, thereby increasing cognitive load and processing time for learners. No significant group differences were found in terms of cognitive load. This may be explained by the limited validity of retrospective self-reports~\cite{Sweller2011}, as well as the structured learning environment, which likely reduced extraneous load for all participants.
Similarly, no significant differences were found between the representations in terms of medium-term retention. Both groups maintained their learning gains over time.
Overall, the findings suggest that not all visually salient representations are equally effective for learning. Although Quantum Beads were more visually prominent, the Bloch sphere enabled more efficient task performance. Future research should therefore take a more differentiated look at the alignment between representational features and specific learning goals, and employ more objective methods—such as eye tracking—to assess cognitive processing more precisely.

\begin{acknowledgments}
\end{acknowledgments}

\appendix

\section{Appendixes}
\subsection{Task Performance for each Task}{\label{app:subsec}}

The following figures show the mean efficiency scores and standard errors (SE) for each task performance item, separated by group. We see a clear difference between the groups in terms of quantum state and quantum measurement, which is also evident in amplitude and superposition. It appears that the students using the Bloch sphere are more efficient in all of the tasks mentioned. For the \textit{phase} task, no significant difference was found between the representations.

\begin{figure}[H]
  \centering
  \includegraphics[width=0.3\textwidth]{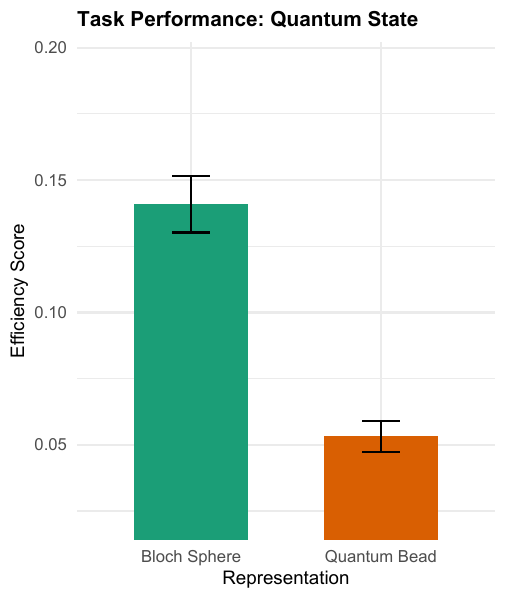}
  \caption{Quantum State}
\end{figure}

\begin{figure}[H]
  \centering
  \includegraphics[width=0.3\textwidth]{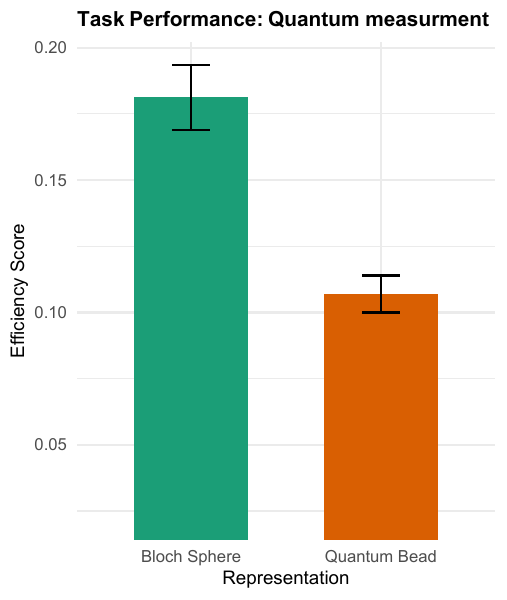}
  \caption{Quantum Measurement}
\end{figure}

\begin{figure}[H]
  \centering
  \includegraphics[width=0.3\textwidth]{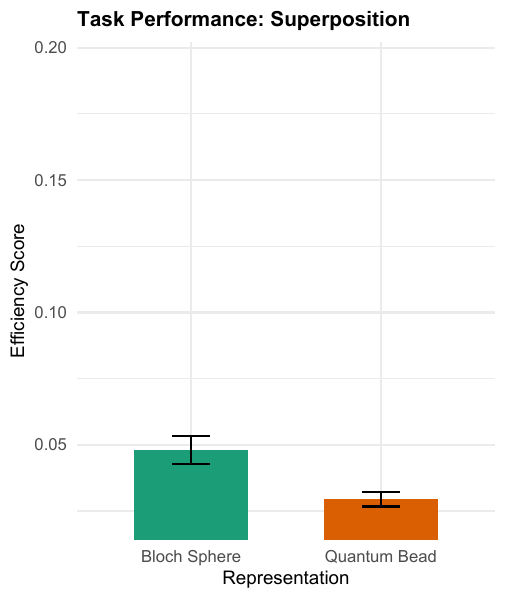}
  \caption{Superposition}
\end{figure}

\begin{figure}[H]
  \centering
  \includegraphics[width=0.3\textwidth]{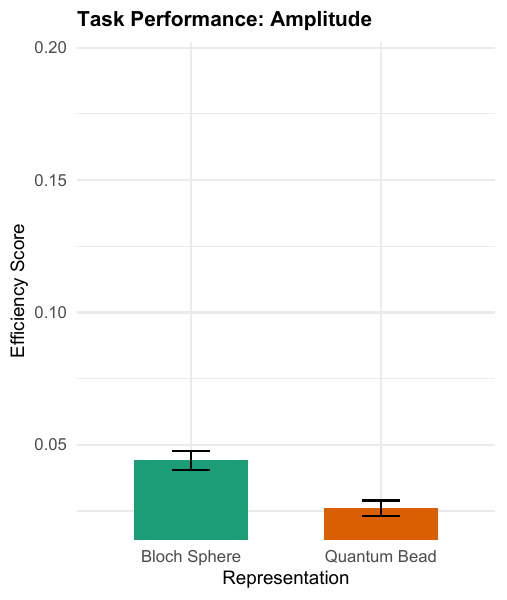}
  \caption{Amplitude}
\end{figure}

\begin{figure}[H]
  \centering
  \includegraphics[width=0.3\textwidth]{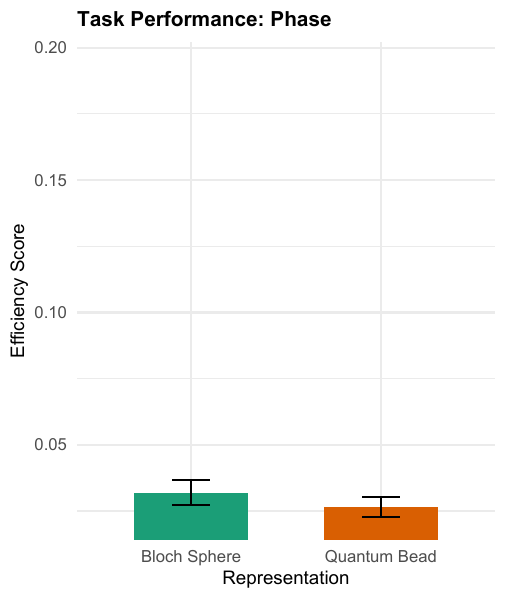}
  \caption{Phase}
\end{figure}

\section{ACKNOWLEDGMENTS}
We would like to thank Silke Stähler-Schöpf, head of the PhotonLab student laboratory at the Max Planck Institute of Quantum Optics (MPQ), for her valuable contributions to this article. We would also like to thank all participants who contributed to this study.

\bibliographystyle{apsrev4-2}
\bibliography{aipsamp}

\section{List of acronyms}
\begin{acronym}
\acro{anova}[ANOVA]{Analysis of variance}
\acro{clt}[CLT]{Cognitive Load Theory}
\acro{ctml}[CTML]{Cognitive Theory of Multimedia Learning}
\acro{deft}[DeFT]{Design, Functions, and Tasks}
\acro{mer}[MERs]{Multiple external representations}
\acro{qp}[QP]{Quantum physics}
\acro{qt}[QT]{Quantum technologies}
\acro{ecl}[ECL]{Extraneous cognitive load}
\acro{icl}[ICL]{Intrinsic cognitive load}
\acro{gcl}[GCL]{Germane cognitive load}
\acro{et}[ET]{Eye tracking}
\acro{sd}[SD]{standard deviation}
\end{acronym}

\end{document}